\documentclass{elsart}

\usepackage{amssymb}
\usepackage{graphicx}

\begin{document}

\begin{frontmatter}




    \vspace{-1.4cm}

\title{Status of IceCube in 2005}
\author{Albrecht Karle\corauthref{cor1}}
\corauth[cor1]{Department of Physics, University of Wisconsin, Madison, 
  U.S.A.}
\author{for the IceCube Collaboration \corauthref{cor2}}
\corauth[cor2]{The full list of authors is available at
http:/icecube.wisc.edu/pub\_and\_doc/conferences/VLVnT2/}

\begin{abstract}
IceCube is a kilometer scale neutrino observatory now in construction
at the South Pole.  The construction started in January 2005 with the
deployment of 76 sensors on the first string and four surface detector
stations.  Nine strings and 32 surface detectors are in operation 
since February 2006. 
The data based on calibration measurements, muons and
artificial light flashes are consistent with performance expectations.
This report focuses on design, construction experience and first data
from the sensors deployed in January 2005.
\end{abstract}

%
\end{frontmatter}


\section{Introduction}

The IceCube neutrino observatory is a kilometer scale neutrino
telescope currently in construction at the South Pole. 
The
feasibility of a neutrino telescope in ice has been demonstrated by
the successful installation and operation of AMANDA. The current
AMANDA array will be surrounded by the IceCube array.  IceCube is
designed to detect astrophysical neutrino fluxes at energies from a
few 100\,GeV up to highest energies of $10^{9}$¥GeV \cite{performance},
\cite{pdd}. 
This report provides an outline of the current construction
status, first experiences with construction and first results and
conclusions from data obtained from sensors now in the ice.

\vspace{-0.2cm}

\section{Design of instrument}

\begin{figure}[htb]
\centering
 \includegraphics[width=0.8\textwidth]{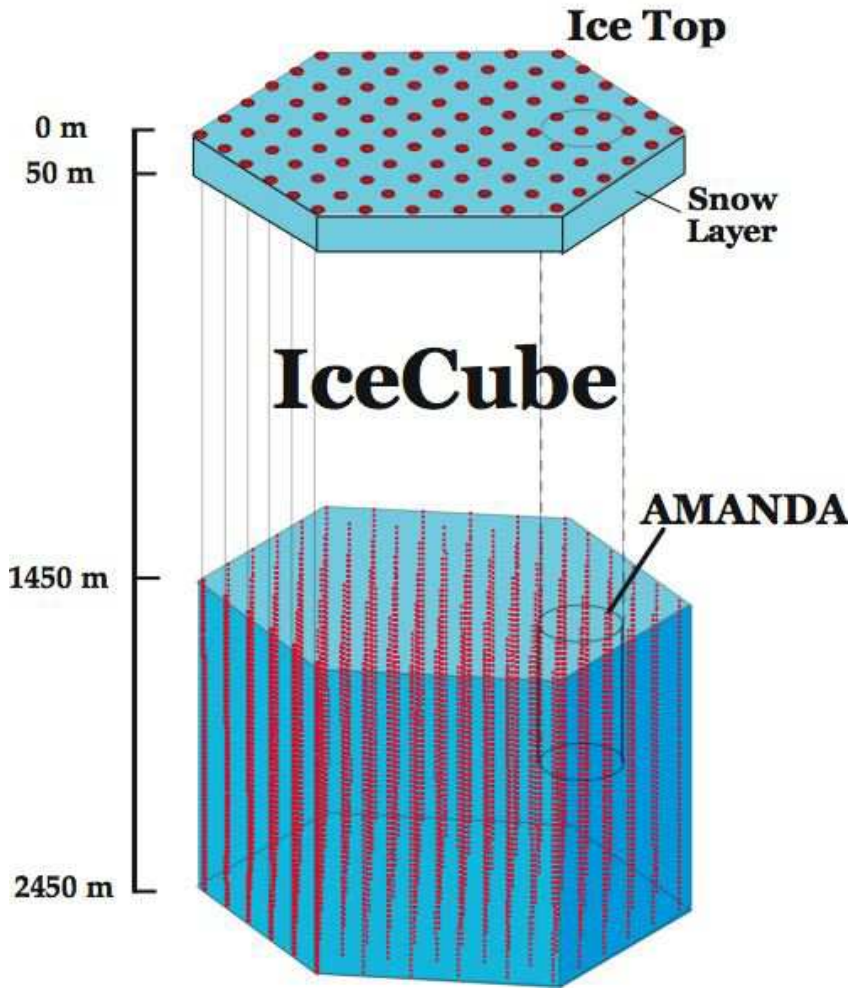}
\caption{Schematic view of the IceCube array consisting of 80 strings 
with 60 sensors on each string.  The surface detector 
IceTop consists of 160 tanks, of which 
two are associated with each string.}
\label{icecube}
\end{figure}

The IceCube neutrino observatory at the South Pole will consist of
4800 optical sensors - digital optical modules (DOMs), installed on 80
strings between the depths of 1450\,m and 2450\,m in the Antarctic
Ice, and 320 sensors deployed in 160 IceTop tanks on the ice surface 
in pairs 
directly above the strings.  Each sensor consists of a 
photomultiplier tube, connected to a waveform-recording data
acquisition circuit capable of resolving pulses with nanosecond
precision and having a dynamic range of at least 250 photoelectrons
per 10\,ns.  
A total of 76 sensors were installed in January 2005,  
60 on the first IceCube string and 16 in the 
first 8 IceTop tanks.  
The table summarizes the construction status as of February 2006.

\begin{tabular}{|c|c|c|c|}
\hline
    Year    &  Strings &   Tanks    &   Sensors   \\
\hline
    Deployed in 2004/05    &   1     &   8    &    76     \\
    Deployed in 2005/06   &   8     &   24   &    528   \\
\hline
    Configuration in 2006   &   9 &  32   &   604   \\
\hline
\end{tabular}
\\

Following the AMANDA concept, it was a design goal to avoid single
point failures in the ice, which is not accessible once frozen.  30
twisted pair copper cables packaged in 15 twisted quads are used to
provide power and communication to 60 sensors.  A string consists of
the following major configuration items: a cable from the counting
house to the string location, a cable from the surface to 2450\,m depth,
and 60 optical sensors.  The main cable is about 2500\,m long and
42\,mm in diameter with a weight of 6\,t. 
Three additional quads provide communication to pressure sensors 
and local communication between optical sensors.
Every 17\,m there is a mounting structure (Yalegrip) 
to allow a quick mechanical attachment of the sensors.
To reduce the amount of cable, two sensors are 
operated on the same wire pair, one terminated and 
one unterminated.  
Neighboring sensors are 
connected to enable fast local coincidence triggering
in the ice.  Each sensor has a direct connection to the 
data acquisition computers in the central counting house. 

\begin{figure}[htb]
\centering
 \includegraphics[width=0.8\textwidth]{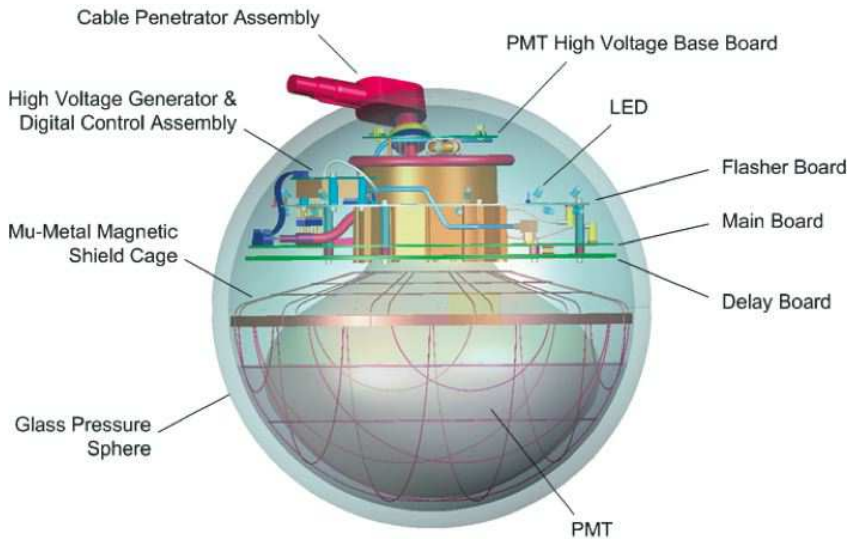}
\hspace*{\fill}
\caption{Schematic view of a Digital Optical Module.}
\label{DOM}
\end{figure}



A schematic view of an optical sensor is shown in Fig. \ref{DOM}.  
An optical sensor consists of a 25\,cm diameter 
photomultiplier tube (PMT) embedded in a glass pressure vessel of 
32.5 cm diameter. 
The HAMAMATSU R7081-02 PMT has ten dynodes 
allowing operation at a gain of at least $5 \cdot 10^{7}$.  
The typical gain in operation 
is  $0.5 - 1.0\cdot 10^{7}$ providing a single photoelectron 
amplitude of about 10\,mV.   
The PMT signal 
is amplified by 3 different gains (x0.25, x2, x16)  
to extend the dynamic range.
The signals are digitized by a fast analog transient waveform recorder 
(ATWD)  and by a flash ADC (40 MHz).  The ATWD uses a 128 sample switched 
capacitor array which is operated at a sampling rate of 3.3 ns/sample.
The linear dynamic range of the sensor is 400 photoelectrons in
15\,ns and the integrated dynamic range is of more than 
5,000 photoelectrons in 2 $\mu$s.
The PMT analog pulse is delayed by 75\,ns on a separate circuit board
to account for the time needed to make the trigger decision
and initiate the ATWD for readout.
The digital electronics on the main board are based on a 
400k-gate field programmable gate array (FPGA) which contains a 
32-bit 
CPU, 8\,MB of flash storage, and 32\,MB of RAM. 
A small communications program stored in ROM 
allows communication to be established with 
the surface computer system and then to 
download updated software.

The flasher board is an optical beacon integrated in each DOM. 
It contains a total of 12 LEDs; 6 pointing horizontally 
and 6 pointing 48 degrees upward.  The LEDs can be fired individually 
or as a group and the amplitude can be adjusted over a wide range
 up to a brightness of $10^{10}$ photons/pulse at a wavelength of about 
410\,nm. 
The flasherboard allows for a variety of calibration functions, 
for example timing and geometry verification. 

The high voltage is generated by an  HV generator, 
which is located on a separate board.  The base is 
a simple resistor chain with appropriate capacitors.  
All electronics of the DOM were designed for high reliability 
and stress testing was performed to screen for high quality. 
Low power consumption was a design goal, the power of a single DOM being 
3\,W in normal operation.

\vspace{-0.2cm}

\section{Data acquisition and online data processing}

All photomultiplier pulses are to be sent to the surface, but 
a local coincidence (LC) trigger scheme is used 
to apply data compression for isolated hits.
Isolated hits are mostly noise pulses. 
For the LC trigger they are defined as pulses for 
which no signal was recorded 
in the neighboring sensors within $\approx\pm800$\,ns. 
All other hits are being transmitted with full 
waveform information. 
The data acquisition system in 2005 is based on 
the system that is used for the final acceptance 
testing of the sensors during and after production. 
A full data acquisition with increased functionality 
will be used in 2006.

The detector electronics and software are designed 
to require minimal maintenance at the remote location.
For example the time calibration system, a critical 
part of a neutrino telescope, is designed to be a
self calibrating, integral part of the readout system 
(in contrast to the AMANDA detector, which required 
manual calibration of all analog detector channels.)
The strings are calibrated as soon as they are frozen-in 
allowing for gradual commissioning of the instrument. 
The collaboration plans to do the first science 
with the array of 9 strings in 2006.  

The full detector will trigger at a rate of about 1.5\,kHz
from down-going muons.  This data set will be reduced 
by at least an order of magnitude, 
small enough for transmission by satellite.  This is done by an online computer 
farm, which will filter events, compress the data volume 
of the muon background and 
select all potentially interesting events for satellite transmission.
A complete copy of the full, unfiltered data set will 
be written on tape for later access.  

\vspace{-0.2cm}
 \begin{figure}[htb]
     \vspace{-0.2cm}
 \centering
  \includegraphics[width=0.8\textwidth]{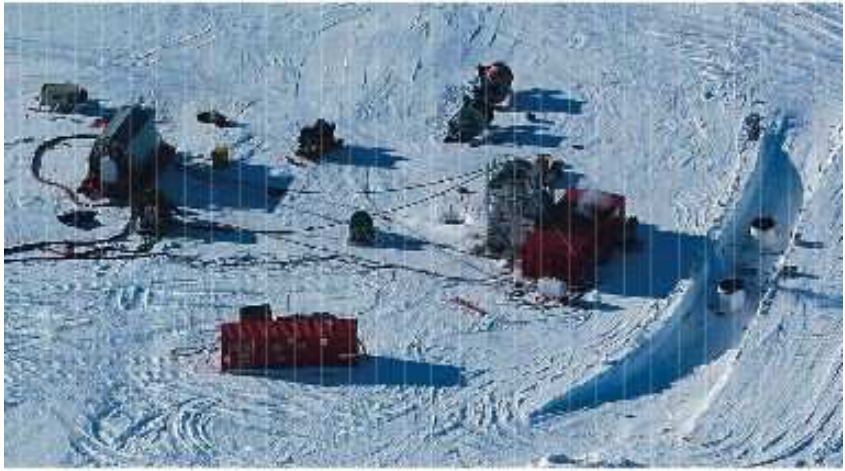}
\vspace{-0.4cm}
 \caption{Aerial view of the tower operations site for the first 
 IceCube string 29 in January 2005: Hose reel on left 
 (weight: 40t),  drill tower with associated heated
 drilling and deployment structure on center-right.  
 The excavation for two IceTop Cherenkov tanks 
 is visible on the right.}
 \label{towersite}
 \end{figure}

\vspace{-0.3cm}

\section{Construction}

  \begin{figure}[htb]
 \centering
  \includegraphics[width=0.8\textwidth]{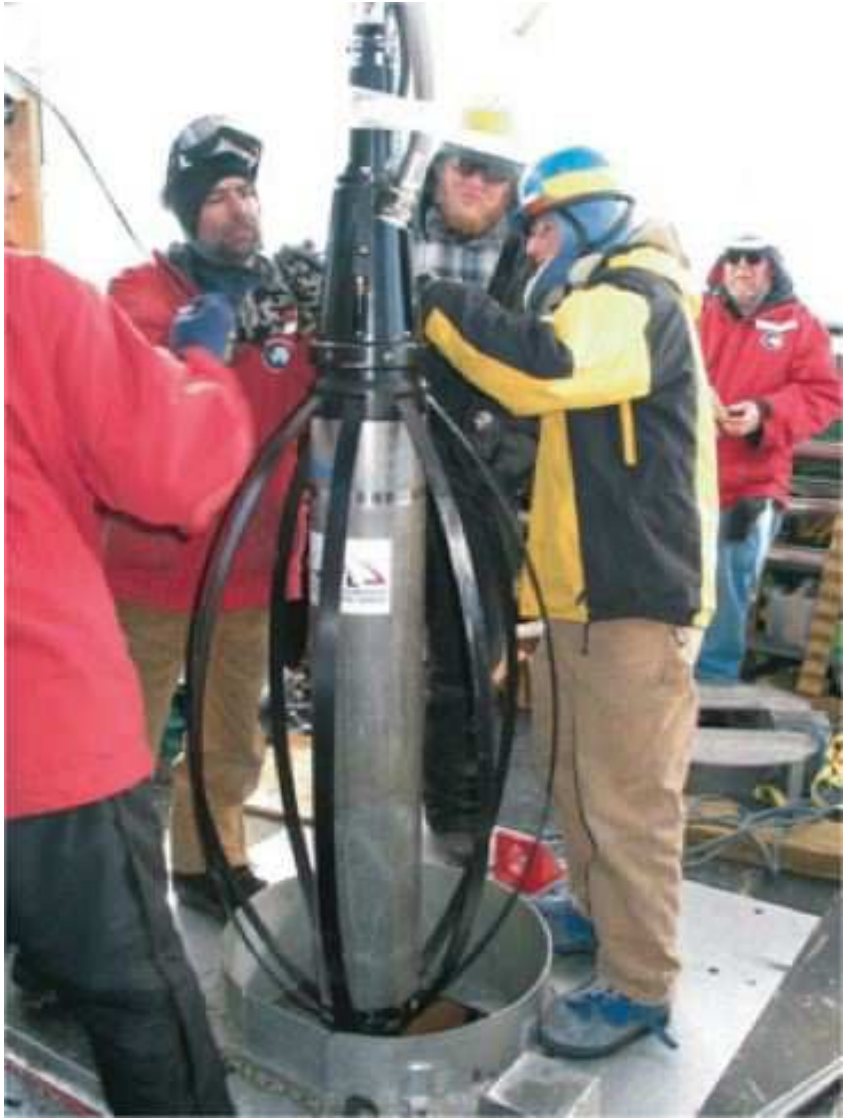}
 \caption{IceCube drillers are rigging the drill moments before 
 it descends to a depth of 2500\,m.}
 \label{drillers}
 \end{figure}

 \begin{figure}[htb]
 \centering
  \includegraphics[width=0.8\textwidth]{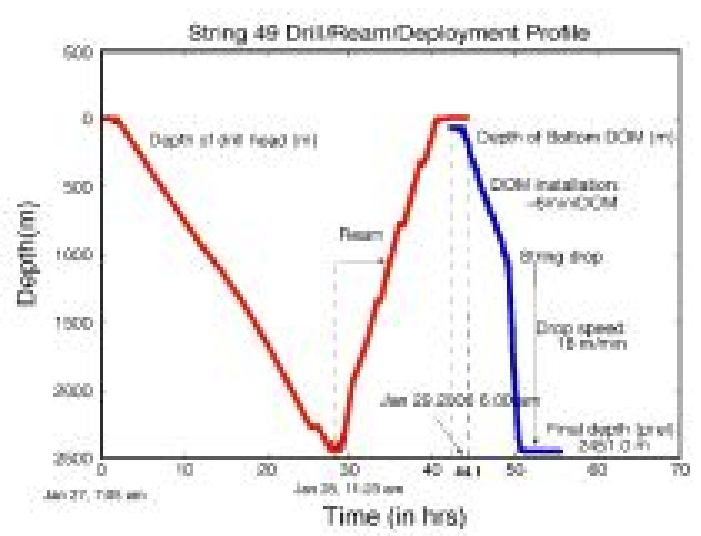}
 \caption{Depth versus time profile for the drillhead during drilling 
 of hole 49.  Also shown is the string installation of the bottom 
 end of the string.}
\label{timeline}
 \end{figure}

The detector is constructed by drilling holes in the ice using a 
hot-water drill.  The first operation of the new enhanced hot water 
drill (EHWD) in January 2005 was a challenging task.  
The drill system consists of numerous pump and heating systems, hoses, a drill 
tower and a complex control system.  While drilling the first hole it 
successfully provided a thermal power of about 5\,MW, consistent with 
the design specification and a factor two more powerful 
than the drill used for AMANDA. 
The drill is designed to drill holes to a depth of 2500\,m 
in a period of less than 
35 hours, excluding the time needed for rigging. 
In comparison, the drilling for AMANDA required typically 90\,h
to reach a depth of 2000\,m. 
Fig. \ref{towersite} shows the drill tower 
with several reels during drilling of the first hole. 
Fig. \ref{drillers} shows the rigging of the drill head
moments before it descends to its 2500\,m deep round trip. 
Fig. \ref{timeline} shows the depth of the drill head 
and the string as a function 
of time for the first hole/string. 
The completed hole is of $\approx$60\,cm diameter and filled with 
water up to about 50\,m below the snow layer. 
The water filled hole freezes within about a week and it is 
designed large enough to allow 24\,h for the string installation.    
The string is then deployed into the water-filled 
hole within less than 12 hours. 

%

This method has been pioneered and developed by AMANDA. IceCube will
use this improved drill and deployment equipment to deploy up to 16
strings during a South Pole construction season, which lasts from
November to mid-February.  

\vspace{-0.2cm}

\section{Engineering data - first measurements} 

The collaboration reported first results, which show that the detector 
system  works as expected \cite{icecube-icrc}.  
All of the 76 sensors were found to communicate and produce 
excellent data. 
A variety of data was taken, including cosmic ray muon data,  
neutrino candidates, light flasher events,  coincident muon events 
with AMANDA and coincident airshower events with IceTop. 
All of these data produced results consistent with 
expectations.

\begin{figure}[htb]
\centering
 \includegraphics[width=0.8\textwidth]{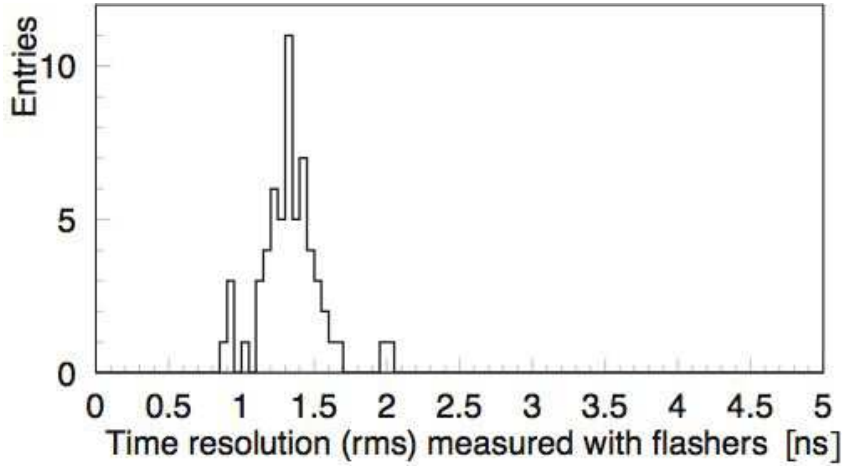}
\caption{The time difference of a flasher pulse
is measured between two optical sensors at 
17\,m and 34\,m distance, respectively.  The time difference
for all sensors constitutes a system level time resolution
verification.}
\label{flasher}
\end{figure}

\begin{figure}[htb]
\centering
 \includegraphics[width=0.8\textwidth]{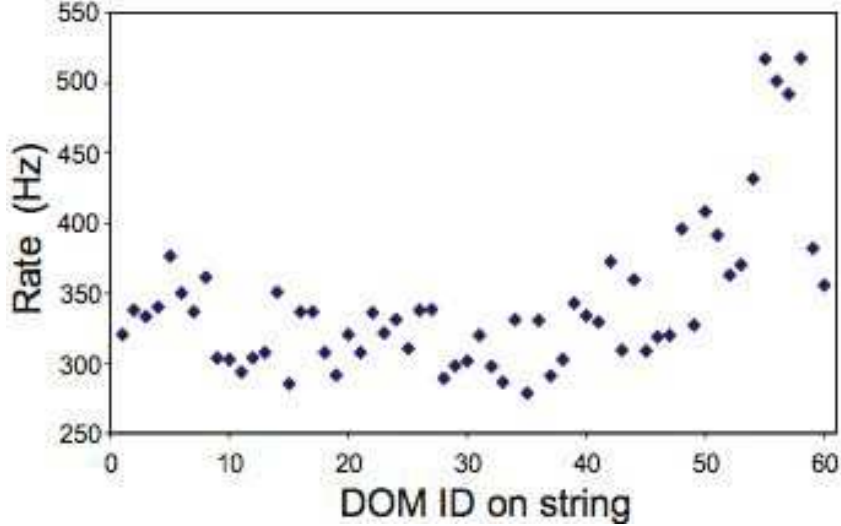}
\caption{Observed single photoelectron noise rates on the first 
IceCube string. The DOM ID from 1 to 60 on the string describes the  
position of the sensor at a depth from 1450\,m to 2450\,m.
The measurement was made with an intentional deadtime of 51\,$\mu$s.}
\label{noise}
\end{figure}


An accurate time calibration system is very important for the
experiment. 
All sensors have precise local quartz oscillators.  
The measurement of the signal transit time through the cable 
is a prerequisite for the determination of the absolute time  of the 
local clocks. This is done by sending a pulse to the sensor,
which is recorded by the DOM and then sent back and recorded again at 
the surface.  Once the cable transit time 
is known the DOMs can be synchronized to GPS time.
The local clocks are synchronized 
every  5 seconds to the central GPS clock at a precision of less 
than 3\,ns.  
LED flashers are used to demonstrate relative timing and geometry. 
Fig. \ref{flasher} shows the time resolution 
of sensors as measured with light flashers. 
The time difference of bright pulses are measured with 
the two adjacent sensors.  
The average rms of less than 1.5\,ns 
is proof that the relative timing of sensors is stable.

Another first measurement, which was awaited with much interest 
was the single photoelectron noise rate of the sensors in situ. 
We know from AMANDA that the noise rates of photomultipliers in ice 
are relatively low, of order 1\,kHz.  There is no known background 
from the ice other than photons generated by cosmic particles. 
The noise background is dominated by light generated in 
the glass of the pressure housing and the glass of the 
photomultiplier.  
For IceCube further efforts had been made to
reduce the number of photons generated by radioactivity in the 
sensors, 
and hence the DOM dark rates.
Once the string was frozen and the temperatures approached 
equilibrium the noise of the sensors settled at a rate of
$\approx$700\,Hz when measured without after-pulse suppression.
The rate is $\approx$350\,Hz as shown in Fig. \ref{noise} 
when measured with after-pulse suppression 
of 51\,$\mu$s.  The first value is relevant for the data rates. 
The latter value 
exceeds earlier expectations. It is important for the 
sensitivity of IceCube 
to detect the low energy neutrino pulse associated with supernova 
explosions. 
Due to improved glass quality the noise rate of the IceCube sensors 
is substantially smaller than for AMANDA, while the optical sensitivity 
is about 1.4 times larger.  

\begin{figure}[htb]
\centering
 \includegraphics[width=0.45\textwidth]{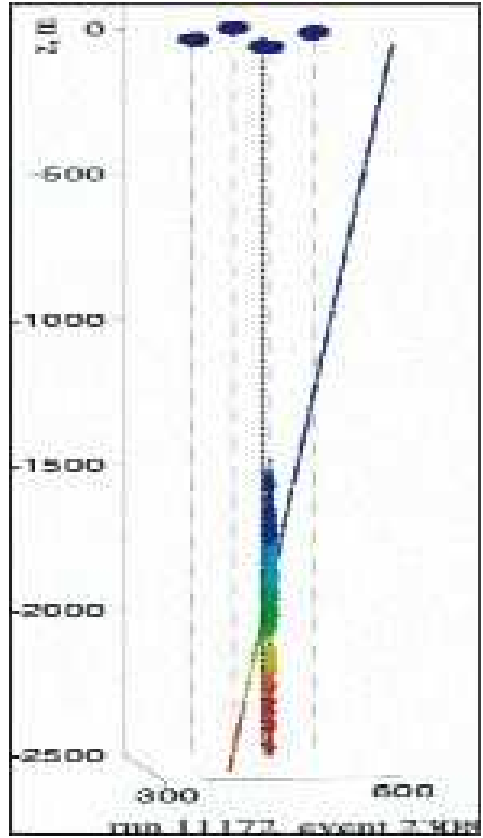}
\caption{Event display of an air shower detected in the surface detector stations 
with a coincident muon observed in string 29.}
\label{event2}
\end{figure}

\begin{figure}[htb]
\centering
 \includegraphics[width=0.7\textwidth]{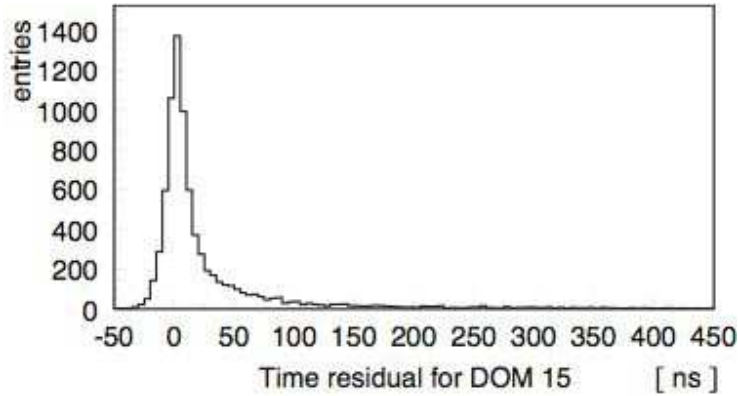}
\caption{ Time delays of measured photons with respect to 
the nominal muon Cherenkov time.  The time residuals
are a verification of the timing system and the detector response 
to muons.}
\label{timeres}
\end{figure}


Muon tracks have been used to test the timing of individual sensors. 
A system level verification of the
time resolution of a sensor can be done by comparing 
the time response of that sensor
with the fitted Cherenkov time based 
on all other sensors on the single string.  An example of an event is 
shown in Fig. 
\ref{event2}. The event has been recorded in eight ice Cherenkov 
detectors at the surface and in most of the sensors on the deep 
string. In this event detector system records an air shower over 
a distance of 2500\,m.  
Fig. \ref{timeres} shows the time delay distribution of 
a sensor with respect to the nominal Cherenkov cone arrival time. 
The distribution is centered around zero with some photons delayed 
due to scattering.

Muon zenith angle distributions taken with the first string showed 
agreements with simulations.
Cosmic ray muons are not only considered as background
but used for several purposes in IceCube. 
While the optical sensors are pointing downward, the sensitivity 
for downward going muons is approximately as good as for upward 
going muons.
Muons can be used to verify the geometry and  
and the detector acceptance using the high 
statistics of downgoing muons. 
Muons may also be used to verify the IceCube absolute pointing 
resolution using the cosmic ray shadow
of the Moon, as has been done with other surface airshower 
and with underground detectors. 
We also plan to do cosmic ray air shower physics using 
deep ice muon data in coincidence
with the surface airshower data taken with the 
detector system IceTop, which is an important part of 
the IceCube observatory. 

\vspace{-0.3cm}

\section{Performance of the complete array}

\begin{figure}[htb]
\centering
\vspace{-0.4cm}
 \includegraphics[width=0.8\textwidth]{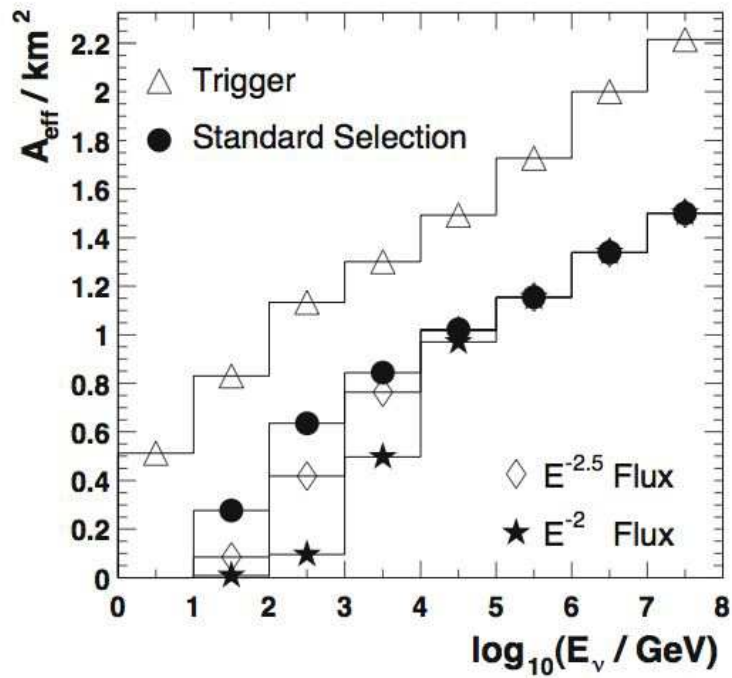}
\vspace{-0.4cm}
 \caption{The effective area of IceCube for muons is shown versus 
muon energy.}
\label{effarea}
\end{figure}

Detailed performance studies were performed using MonteCarlo
simulations \cite{performance}, \cite{pdd}. 
The reconstruction algorithms used for AMANDA yield an angular 
resolution of $0.7^{\circ}$ at 10\,TeV. 
The effective area for muons is about 0.8\,km$^{2}$ for 1 TeV muons
after background rejection and exceeding 1\,km$^{2}$ at higher energies. 
IceCube is expected to trigger at a rate of 1.5\,kHz on cosmic ray muons
and yield about $10^{5}$ atmospheric neutrinos per year.   
The effective area for muons after background rejection \
is shown in Fig. \ref{effarea} \cite{performance}. 
The combined underground and surface detector will allow 
the measurement of air shower parameters in detail beyond 
energies of $10^{9}¥$GeV. 
During the construction phase until 2010 IceCube will 
grow year be year in size and performance providing 
a steady inrease in sensitivity.

\vspace{-0.3cm}

\vspace{-0.3cm}

\section{Conclusions}

The successful deployment of the first IceCube string 
in January 2005 provided the basis for confidence 
in the detector design and the overall  
construction schedule. 
The IceCube collaboration was able to extract 
data from the first 76 sensors that show that the detector works.
Noise rates, timing calibrations and muon distributions 
are consistent with expectations.  
Eight more strings were deployed in the 2005/06 season. 
IceCube is already now, with 9 strings and 16 IceTop stations, 
the largest neutrino detector  taking data 
in the world with a total of 604 optical sensors deployed. 
Moreover it can be jointly operated  with the existing
677 optical sensor AMANDA-II array.   
The combined detector 
has started taking science quality data in 2006.

\vspace{-0.1cm}

\section{Acknoweldgements}
This research was supported by the Deutsche Forschungsgemeinschaft
(DFG); German Ministry for Education and Research; Knut and Alice
Wallenberg Foundation, Sweden; Swedish Research Council; Swedish
Natural Science Research Council; Fund for Scientific Research
(FNRS-FWO), Flanders Institute (IWT), Belgian Federal Office for
Scientific, Technical and Cultural affairs (OSTC), Belgium. UC-Irvine
AENEAS Supercomputer Facility; University of Wisconsin Alumni Research
Foundation; U.S. National Science Foundation, Office of Polar
Programs; U.S. National Science Foundation, Physics Division;
U.S. Department of Energy.

\end{document}